# Potential Neutralizing Antibodies Discovered for Novel Corona Virus Using Machine Learning

*Rishikesh Magar [¥], Prakarsh Yadav [‡] and Amir Barati Farimani [1¥§‡]*

[¥] Department of Mechanical Engineering, Carnegie Mellon University, Pittsburgh, PA 15213

[‡] Department of Biomedical Engineering, Carnegie Mellon University, Pittsburgh, PA 15213

[§] Machine Learning Department, School of Computer Science, Carnegie Mellon University
Pittsburgh, PA 15213

**Abstract**

The fast and untraceable virus mutations take lives of thousands of people before the immune system can produce the inhibitory antibody. Recent outbreak of novel coronavirus infected and killed thousands of people in the world. Rapid methods in finding peptides or antibody sequences that can inhibit the viral epitopes of COVID-19 will save the life of thousands. In this paper, we devised a machine learning (ML) model to predict the possible inhibitory synthetic antibodies for Corona virus. We collected 1933 virus-antibody sequences and their clinical patient neutralization response and trained an ML model to predict the antibody response. Using graph featurization with variety of ML methods, we screened thousands of hypothetical antibody sequences and found 8 stable antibodies that potentially inhibit COVID-19. We combined bioinformatics, structural biology, and Molecular Dynamics (MD) simulations to verify the stability of the candidate antibodies that can inhibit the Corona virus.

---

[1] Corresponding Author, e-mail: barati@cmu.edu, Website: www.baratilab.com

**Introduction**

The biomolecular process for recognition and neutralization of viral particles is through the process of viral antigen presentation and recruitment of appropriate B cells to synthesize the neutralizing antibodies.[1] Theoretically, this process allows the immune system to stop any viral invasion, but this response is slow and often requires days, even weeks before adequate immune response can be achieved.[2,3] This poses a challenging question: Can the process of antibody discovery be accelerated to counter highly infective viral diseases?

With the rapid expansion of available biological data, such as DNA/protein sequences and structures[4], it is now possible to model and predict the complex biological phenomena through machine learning (ML) approaches. Given sufficient training data, ML can be used to learn a mapping between the viral epitope and effectiveness of its complementary antibody.[5] Once such mapping is learnt, it can be used to predict potentially neutralizing antibody for a given viral sequence[6].

ML can essentially learn the complex antigen-antibody interactions faster than human immune system, leading to the generation of synthetic inhibitory antibodies acting as a bridge, which can overcome the latency between viral infection and human immune system response. This bridge can potentially save the life of many especially during an outbreak and pandemic. One such instance is the spread of coronavirus disease (COVID-19)[7].

With incredibly high infectivity and mortality rate, COVID-19 has become a global scare.[8,9] To compound the problem, there are no proven therapeutics to aid the suffering patients[2,8,10–14]. Only viable treatment at the moment is symptomatic and there is a desperate need for developing therapeutics to counter COVID-19. Recently, the proteomics sequences of 'WH-Human 1' coronavirus became available through Metagenomic RNA sequencing of a patient in Wuhan.[4,15] WH-Human 1 is 89.1% similar to a group of SARS-like coronaviruses.[4] With this sequence available, it is possible to find potential inhibitory antibodies by scanning thousands of antibody sequences and discovering the neutralizing ones[16–18]. However, this requires very expensive and time-consuming experimentation to discover the inhibitory responses to Corona virus in a timely manner. In addition, computational and physics-based models require the bound crystal structure of antibody-antigen complex, however; only a few of these structures have become available.[19,20,21,22] In the case of COVID-19, the bound antigen-antibody crystal structure is not available to-date[23,24]. Given this challenge and the fact that ML models require a large amount of data, the ML approach should rely on the sequences of the antibody-antigen rather than the crystal structures[25].

In this paper, we have collected a dataset comprised of antibody-antigen sequences of variety of viruses including HIV, Influenza, Dengue, SARS, Ebola, Hepatitis, etc. with their patient clinical/biochemical $IC_{50}$ data. Using this dataset (we call it VirusNet), we trained and benchmarked different shallow and deep ML models and selected the best performing model. Based on SARS 2006 neutralizing antibody scaffold[26], we created thousands of potential antibody candidates by mutation and screened them with our best performing ML model. Finally, molecular dynamics (MD) simulations were performed on the neutralizing candidates to check their structural stability. We predict 8 structures that were stable over the course of simulation and are potential neutralizing antibodies for COVID-19.

In addition, we interpreted the ML method to understand what alterations in the sequence of binding region of the antibody would most effectively counter the viral mutation(s) and restore the ability of the antibody to bind to the virus[27]. This information is critical in terms of antibody design and engineering and reducing the dimension of combinatoric mutations needed to find a neutralizing antibody.

# Methods

1. Dataset

The majority of the data in the training set is composed of HIV antibody-antigen complex (1887 samples). Most of the samples for the HIV training set were obtained from the Compile, Analyze and Tally NAb panels (CATNAP) database from the Los Alamos National Laboratory (LANL)[28,29]. From CATNAP, data was collected for monoclonal antibodies, 2F5, 4E10 and 10E8, which bind with GP41[30–32]. Using CATNAP's functionality for identifying epitope alignment, we selected FASTA sequence of the antigen corresponding to the site of alignment, in the antibody. We generated a dataset with 1831 training examples comprising of antibodies – antigen sequences and their corresponding $IC_{50}$ values. The CATNAP output is comprised of site of antigen sequence alignment for each of the antibodies with respect to 2F5, 4E10 and 10E8. Using the co-crystalized structure of (2F5-ELDKWAS) in (PDB:1TJG)[30] as template, the antibody fragment that comes in contact with the antigen was found by considering amino acids within 7Å of the antigen in the co-crystallized structure.

To make the dataset more diverse and train a more robust ML model, we included more available antibody-antigen sequences and their neutralization potential. To do this, we compiled the sequences of Influenza, Dengue, Ebola, SARS, Hepatitis, etc.[26,33–86] by searching the keywords of "virus, antibody" on RCSB server[87] and selected the neutralizing complex by reading their corresponding publications. Furthermore, for each neutralizing complex, the contact residues at the interface of antibody and antigen were selected. To select the antigen contact sequences, all amino acids within 5Å of corresponding antibody were chosen. (Supporting Information) To select the antibody contact sequences, all amino acids within 5Å of the antigen were chosen. In total, 102 sequences of antibody-antigen complexes were mined and added to the 1831 samples, resulting in total number of 1933 training samples.

## 2. Graph Featurization and Machine Learning

For effective representation of molecular structure of amino acids, the individual atoms of amino acids of antibody and antigen were treated as undirected graph, where the atoms are nodes and bonds are edges[88]. It has been shown that graph representation is better in transferring the chemistry and topology of molecular structure compared to Extended Connectivity Fingerprints (ECFP)[88,89]. We construct these molecular graphs using RDkit[90]. Embeddings are generated to encode relevant features about the molecular graph[91,92]. These embeddings encode information like the type of atom, valency of an atom, hybridization state, aromaticity etc. First, each antibody and antigen were encoded into separate embeddings and then concatenated into a single embedding for the entire antibody-antigen complex. We then apply mean pooling over the features for this concatenated embedding to ensure dimensional consistency across the training data. The pooled information is then passed to classifier algorithms like XGBoost[93], Random Forest[94], Multilayer perceptron, Support Vector Machine (SVM)[95] and Logistic Regression which then predict whether the antibody is capable of neutralizing the virus.

## 3. Hypothetical Antibody Library Generation

In order to find potential antibody candidates for COVID-19, 2589 different mutant strains of antibody sequences were generated based on the sequence of SARS neutralizing antibodies. The reason we selected these antibodies as initial scaffolds is that the genome of COVID-19 [4] is 79.8% identical to "Tor2" isolate of SARS (Accession number: AY274119)[96]. Using 4 different antibody variants of SARS (PDB: 2GHW, 3BGF, 6NB6, 2DD8)[26,76,81,86], point mutations were applied to all the amino acids in the binding region of antibody. (see Supporting Information for COVID-19 antigen and antibody interactions) To find out the binding region of these antibodies for sequence generation, all amino acids within 5Å of their respective antigen were chosen. To assess the biological feasibility of these mutant sequences, we scored each mutation by using the BLOSUM62 matrix[97].

4. **Molecular Dynamics Simulations**

To assess the stability of proposed antibody structures, we performed molecular dynamics (MD) simulations of each of antibody structure in a solvated environment[98]. The simulation of solvated antibody was carried out using GROMACS-5.1.4[99–101], and topologies for each antibody were generated according the GROMOS 54a7[102] forcefield. The protein was centered in a box, extending 1 nanometer from surface of the protein. This box was the solvated by the SPC216 model water atoms, pre-equilibrated at 300K. The antibody system in general carried a net positive charge and it was neutralized by the counter ions. Energy minimization was carried out using steepest descent algorithm, while restraining the peptide backbone to remove the steric clashes in atoms and subsequently optimize solvent molecule geometry. The cut-off distance criteria for this minimization were forces less than 100.0 kJ/mol/nm or number of steps exceeding 50,000. This minimized structure was the sent to two rounds of equilibration at 300K. First, an NVT ensemble for 50 picoseconds and a 2-femtosecond time step. Leapfrog dynamics integrator was used with Verlet scheme, neighbor-list was updated every 10 steps. All the ensembles were under Periodic Boundary Conditions and harmonic constraints were applied by the LINCS algorithm[103]; under this scheme the long-range electrostatic interactions were computed by Particle Mesh Ewald (PME) algorithm[104]. Berendsen thermostat was used for temperature coupling and pressure coupling was done using the Parrinello-Rahman barostat[105,106]. The last round of NPT simulation ensures that the simulated system is at physiological temperature and pressure. The system volume was free to change in the NPT ensemble but in fact did not change significantly during the course of the simulation. Following the rounds of equilibration, production run for the system was carried out in NPT and no constraints for a total of 15 nanoseconds, under identical simulation parameters.

**Results and Discussions**

The flowchart of COVID-19 antibody discovery using ML has four major steps (Figure 1): 1. Collecting data and curating the dataset for training set. 2. Featurization, embedding and benchmarking ML models and selecting the best performing one. 3. Hypothetical antibody library generation and ML screening for neutralization and 4. Checking the stability of proposed antibodies. This workflow enables the rapid screening of large space of potential antibodies to neutralize COVID-19. In general, this workflow can be used for high throughput screening of antibodies for any type of virus by only knowing the sequences of antigen epitopes.

To better understand the diversity and similarity of the sequences that were used in the training set, t-Distributed Stochastic Neighbor Embedding (t-SNE) of all different viruses were computed (Figure 2a). t-SNE axes shows the directions of the maximum variance in the dataset, therefore, the dimensionality of the data can be reduced to lower dimensions (here two). HIV antigen shows the most variations on t-SNE components where viruses such as Influenza, Dengue and H1N1 are very close to each other. The neutralizing antibodies were also projected using t-SNE to show the variations in the available neutralizing sequences. (Figure 2b) Unlike antigen variations, antibody sequences are much closer to the center of t-SNE with a few scattered ones. The comparison of Figure 2a and Figure 2b shows that the neutralizing antibodies are not sequence-diverse compared to virus antigens. This difference demonstrates that a large space of potential antibodies can be screened and used for finding novel antibodies. The labels in the dataset are comprised of the neutralization panel data, $IC_{50}$ values for monoclonal antibodies and pseudo-typed viruses. (Figure 2c) The $IC_{50}$ labels were collected from 49 published neutralization studies and were collected from Los Alamos HIV Database (for 1831 samples in our training set). For some cases in CATNAP, personal communication with the authors were made to resolve sequence name ambiguities between different laboratories. For 102 samples of various viruses collected from RCSB server, all of them neutralize their antigen based on biochemical assays. These samples were labeled by setting their $IC_{50}$ to zero. Since classification is performed on the training dataset, $IC_{50} \leq 10$ are set to neutralizing and $IC_{50} > 10$ to non-neutralizing (Figure 2c). To visualize the diversity of the virus types used in the dataset other than HIV, the distribution of 13 more viruses were presented in Figure 2d. Influenza, Dengue, SARS, Ebola and then Hepatitis have relatively larger samples in the dataset.

To benchmark the performance of different ML models on the VirusNet dataset and select the best performing one, XGBoost, Random Forest (RF), Multilayer perceptron (MLP), Support Vector Machines (SVM), and Logistic Regression (LR) were used (Figure 3a). The five-fold cross validation on 80%-20% split, train, and test resulted in best accuracy for XG-Boost model. The performance and ranking of models follow the order of XGBoost (90.57%)> RF (89.18%)>LR (81.17%)> MLP (78.23)> SVM (75.49%). Since the training data is sparse in the case of VirusNet (See Figure 2a and Figure 2b), XGBoost selects the sparse features input by pruning and learning the underlying sparsity patterns. To test the robustness of the XGBoost on completely unseen virus types, for each left-out virus type, the model was trained on all the sequences in the VirusNet except for the left-out. For example, for Influenza, all the sequences and labels of Influenza were removed from the training set and the trained model on the remaining dataset were tested on all the Influenzas's sequences and consequently the classification accuracy were reported (Figure 3b). The accuracies for the out of class test is as follows: Influenza (84.61%), Dengue (100%), Ebola (75%), Hepatitis (75%), SARS (100%). The out of class results demonstrate that our model is capable of generalizing the prediction to a completely novel virus epitope. Since COVID-19 is completely a new virus, we can conclude that our model prediction performance will be accurate. The fact that our model prediction is 100% for SARS out of class test demonstrate its capability of effectively predicting the antibodies for COVID-19 which is from SARS family.

Next, using the best performing model (XGBoost), all the hypothetical candidates in the library were evaluated for neutralization. Out of all the candidates, some of them are invalid mutations screened using BLOSUM62 matrix[97] (Figure 3c). 18 final candidates are both valid mutations and can neutralize COVID-19 with high confidence probability of 0.9895 (shown with green color in Figure 3c).

Interpretability of the ML models is very important in both explaining the underlying biological and chemical understanding of neutralization and providing design guidelines for antibody engineering. One of the significant advantages of ensemble methods such as XGBoost is their interpretability. By taking advantage of this property, the important features that are giving rise to neutralization were ranked and scored (Figure 3d). The input features to the model contains atomic level attributes such as atom type, valency, hybridization, etc. To collectively translate the important atomic features into important amino acid features, the scores of amino acids with unique atomic features were summed up and ranked (Figure 3d). Some of the atomic features were common among all the amino acids (e.g. Carbon, implicit valency, Oxygen, etc.) therefore; we ignored them. However, some of the other features like aromaticity or having Sulfur are unique and we considered those in amino acid features. Based on the unique features appeared in ranking, Methionine (M) is the most important one. (Figure 3d)

The predicted sequences from the ML model were then used to model the novel structure of potentially neutralizing antibodies. The predicted sequences were projected onto their progenitor antibody and the changes in amino acid sequence were modelled as follows: Simple point mutations were introduced by modifying the target amino acid using Coot [107,108](Crystallographic Object-Oriented Toolkit). Coot environment allowed us to predict the stereochemical effect of each point mutation and appropriately compensate for it. Using such an approach, we were able to accurately model the putative structures of the antibodies. The modelled structures were then passed to MD simulations for stability check.

To check the stability of predicted structures energetically, 20 MD simulations (18 point mutations+2 wildtype (WT)) in total were performed (Figure 4a). Structures with low Root Mean Square Deviation (RMSD) and low contact distance are in a stable conformation, whereas structures with high RMSD and high contact distances are in an unstable conformation. RMSD and contact distance for WT structures have lower values, demonstrating stability, therefore; the contact distances versus RMSD is a good indicator of the stability of a protein over the course of a simulation (Figure 4b).

Once mutation introduced in the crystallographic structure, it will cause it to deviate from WT structure's RMSD and contact distance. We performed simulations for all the 18 point-mutant structures and their mean contact distance versus RMSD[109,110] were computed for their respective trajectories (Figure 4b) (see Supporting Information). Based on the two WT structures mean RMSD and contact distances, we selected the mutations which have mean contact distance and RMSD values less than 0.488 nm and 0.35 nm, respectively. (the shaded triangle region in Figure 4b). Candidates with higher values of mean RMSD and contact distances are unstable and will potentially fail to neutralize the COVID-19.

In order to be more comprehensive, we created co-mutations out of 5 stable point mutations (C3, C7, C14, C17, C18, see Table S1 in Supporting Information for the list of all 18 candidates). This resulted in 5 new structures (Co1, Co2, Co3, Co4, Co5 in Table S2) that were screened using XGBoost for neutralization. Among all 5 co-mutations, Co5 did not neutralize. To check the stability of these 4 neutralizing co-mutations, MD simulations were performed and Co1, Co2 and Co4 were found to be stable (Figure 4b). The list of the final 8 stable mutations and co-mutations are tabulated in Table 1 and the PDB structures are available as Supporting Information.

**Conclusion**

We have developed a machine learning model for high throughput screening of synthetic antibodies to discover antibodies that potentially inhibit the COVID-19. Our approach can be widely applied to other viruses where only the sequences of viral coat protein-antibody pairs can be obtained. The ML models were trained on 14 different virus types and achieved over 90% fivefold test accuracy. The out of class prediction is 100% for SARS and 84.61% for Influenza, demonstrating the power of our model for neutralization prediction of antibodies for novel viruses like COVID-19. Using this model, the neutralization of thousands of hypothetical antibodies was predicted, and 18 antibodies were found to be highly efficient in neutralizing COVID-19. Using MD simulations, the stability of predicted antibodies were checked and 8 stable antibodies were found that can neutralize COVID-19. In addition, the interpretation of ML model revealed that mutating to Methionine and Tyrosine is highly efficient in enhancing the affinity of antibodies to COVID-19.

**Acknowledgements**

The authors gratefully acknowledge the use of the supercomputing resource Arjuna provided by the Pittsburgh Supercomputing Center (PSC). This work is supported by Center for Machine Learning in Health (CMLH) at Carnegie Mellon University (https://www.cs.cmu.edu/cmlh-cfp) and start-up fund from Mechanical Engineering Department at CMU. The authors would like to thank Prof. Reeja Jayan for her support and Junhan Li for his help in collecting the data.

**Supporting Information Available**

The RMSD and contact distance plots for all the trajectories versus time, the structure of virus antibody complex and the residues at the contact region, native contacts in antigen-antibody complex, the interaction of COVID-19 epitope with 2GHW antibody, tables for all neutralizing point mutations, co-mutations and their neutralization potentials, the structures of stable antibodies in PDB format, and $IC_{50}$ data interpretation are available online. The VirusNet dataset will be available upon request.

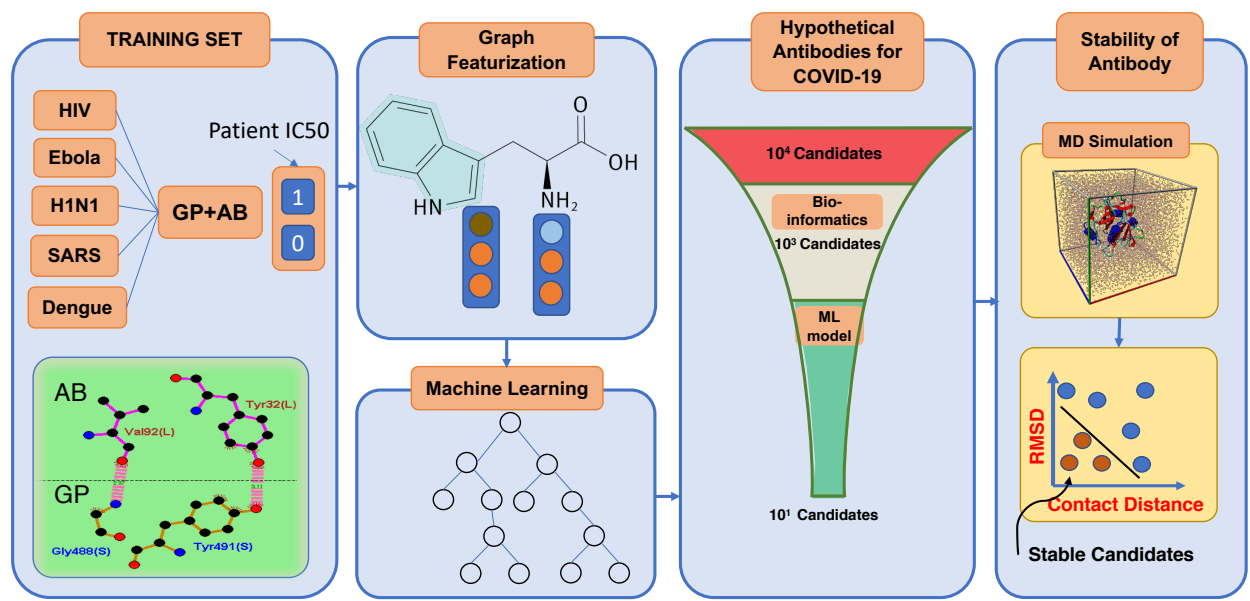

**Figure 1**. Designing antibodies or peptide sequences that can inhibit the COVID-19 virus requires high throughput experimentation of vastly mutated sequences of potential inhibitors. The screening of thousands of available strains of antibodies are prohibitively expensive, and not feasible due to lack of available structures. However, machine learning models can enable the rapid and inexpensive exploration of vast sequence space on the computer in a fraction of seconds. We collected 1933 virus-antibody sequences with clinical patient IC$_{50}$ data. Graph featurization of antibody-antigen sequences creates a unique molecular representation. Using graph representation, we benchmarked and used a variety of shallow and deep learning models and selected XGBoost because of its superior performance and interpretability. We trained our model using a dataset including 1,933 diverse virus epitope and the antibodies. To generate the hypothetical antibody library, we mutated the SARS scaffold antibody of 2006 (PDB:2GHW) and generated thousands of possible candidates. Using the ML model, we classified these sequences and selected the top 18 sequences that will neutralize COVID-19 with high confidence. We used MD simulations to check the stability of the 18 sequences and rank them based on their stability.

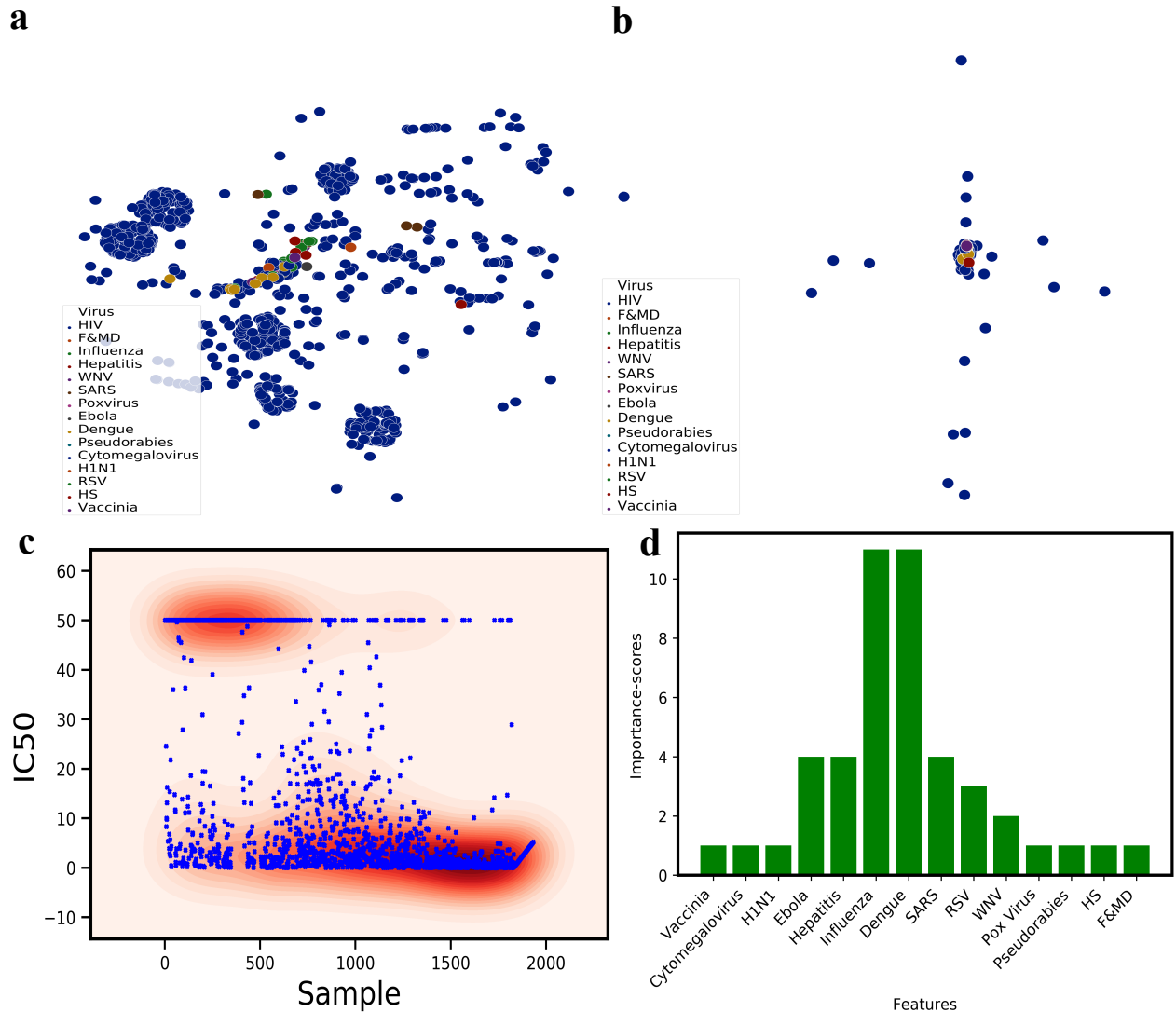

**Figure 2**. **a)** t-Distributed Stochastic Neighbor Embedding (t-SNE) of all the viruses epitopes used in the training dataset, revealing biological similarity and diversity of the sequences used in the dataset. **b)** t-SNE of all the therapeutics antibody sequences used in the training set for variety of different virus types. The majority of the broadly neutralizing antibodies such as 2F5 is clustered at the center of this plot. **c)** Patient clinical IC50 data obtained from various sources and the distribution of the neutralizing ($IC_{50}<10$) and Non- neutralizing ($IC_{50}>10$) samples. **d)** The number of samples for each virus class except HIV. For HIV, we collected 1883 samples. Influenza and Dengue has 10+ samples

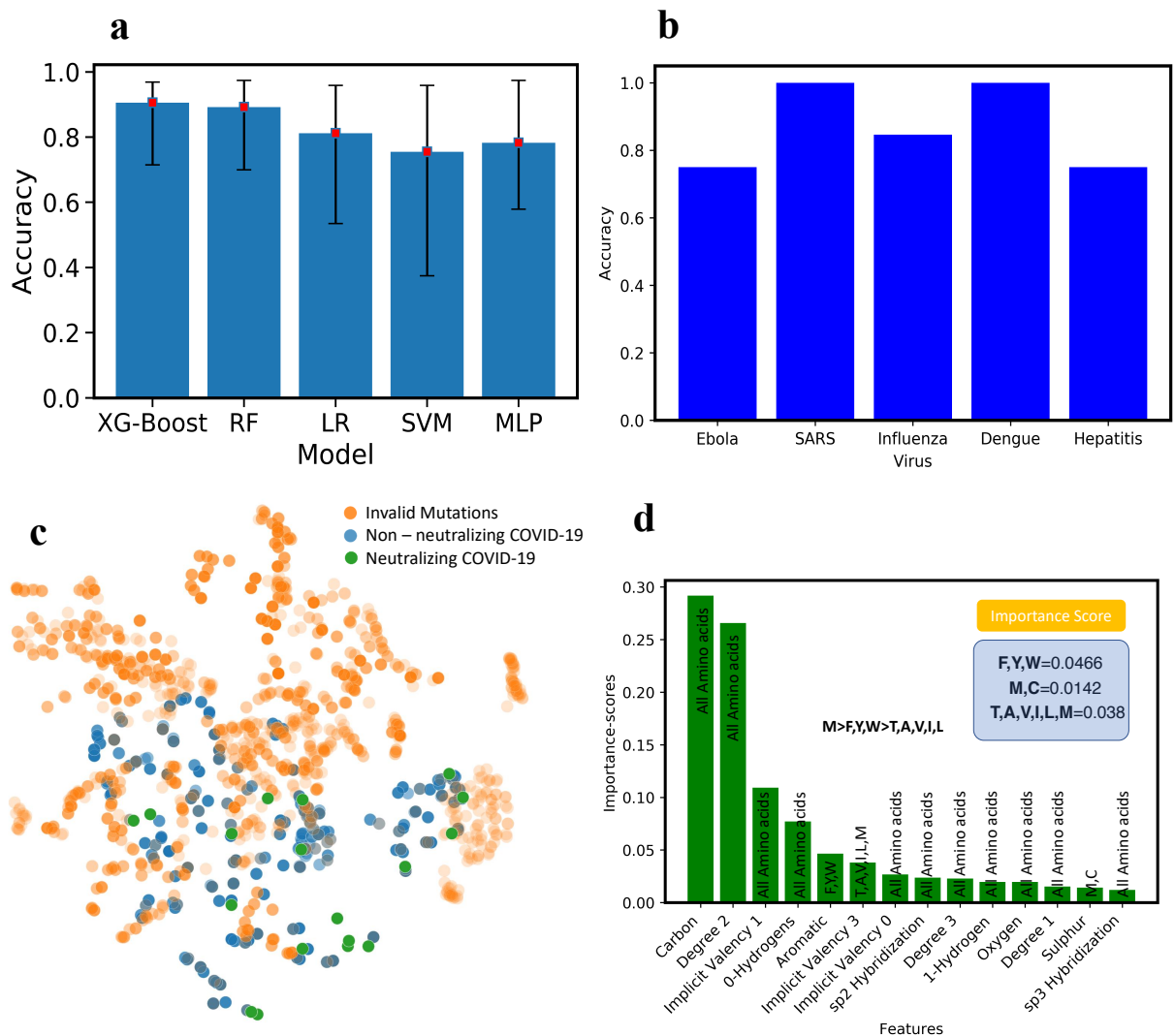

**Figure 3**. **a)** The test accuracy with five-fold cross validation for XG-Boost, Random Forrest (RF), Logistic Regression (LR), Support Vector Machine (SVM) and Deep Learning (Multilayer Perceptron. XGBoost has the highest performance with (90.75%). **b)** Out of training class test accuracy for influenza, Dengue, Ebola, Hepatitis, and SARS. To perform this test, for example for influenza, all the influenza virus-antibody sequences were removed from the training set and the obtained model were tested on all samples of Influenza and the accuracy is reported here. **c)** Blossum validated mutations, non-neutralizing and neutralizing antibody sequences. To achieve more confidence, we set the threshold of prediction probability to 0.9895 in XGBoost and found 18 neutralizing antibody sequences (the green points). **d)** Interpretability of ML model: to understand what mutations are playing the key roles in neutralization, XGBoost feature importance used with ranked atomic level features. Through connecting the atomic features with each of 20 amino acids, M was found to be the most important amino acids in neutralization followed by F, Y, W.

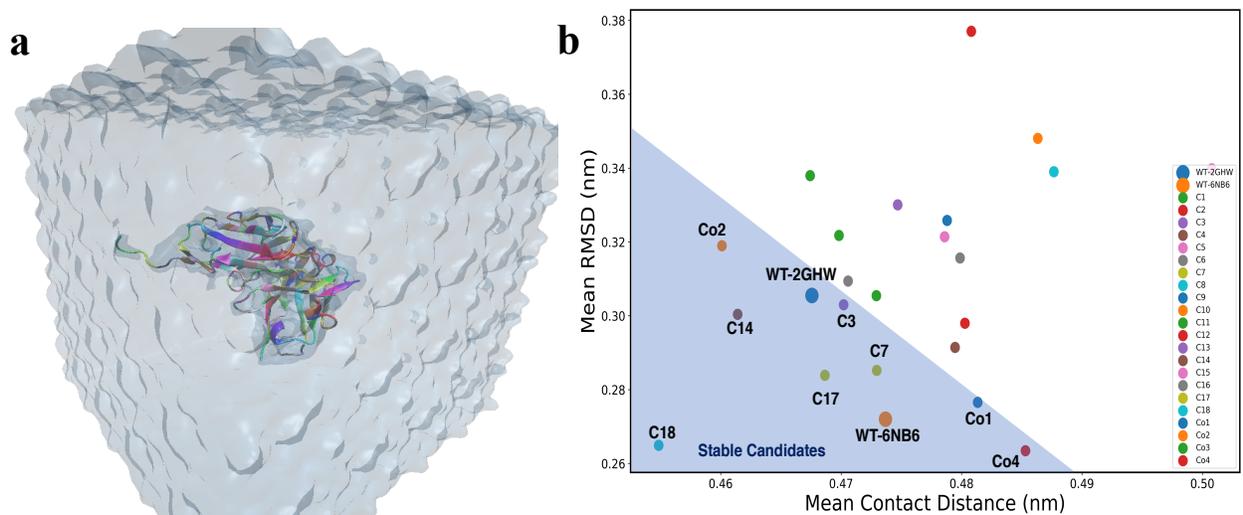

**Figure 4**. **a)** The snapshot of MD simulation of mutated proteins. Each protein is solvated in a box of water and simulated to collect the statistical data on the stability of mutants and co-mutants. **b)** Mean Root Mean Square Deviation (RMSD) versus Mean contact distances for each candidate averaged over the whole trajectory.

| Structure | Mutation |
|---|---|
| C3 | 2GHW–I51M |
| C7 | 2GHW–R150H |
| C14 | 2GHW–T204N |
| C17 | 6NB6–R56H |
| C18 | 6NB6–K58S |
| Co1 | 2GHW-I51M, R150H, T204N |
| Co2 | 2GHW-I51M, R150H |
| Co4 | 6NB6-R56H, K58S |

**Table 1**: The final neutralizing candidates obtained through screening with ML model, MD simulation for stability and Bioinformatics. The detailed list of sequences is available in the Supporting Information.

(27) Zhang, C.; Zheng, W.; Huang, X.; Bell, E. W.; Zhou, X.; Zhang, Y. Protein Structure and Sequence Re-Analysis of 2019-NCoV Genome Does Not Indicate Snakes as Its Intermediate Host or the Unique Similarity between Its Spike Protein Insertions and HIV-. 11.

(28) Yoon, H.; Macke, J.; West, A. P.; Foley, B.; Bjorkman, P. J.; Korber, B.; Yusim, K. CATNAP: A Tool to Compile, Analyze and Tally Neutralizing Antibody Panels. *Nucleic Acids Res.* **2015**, *43* (W1), W213–W219. https://doi.org/10.1093/nar/gkv404.

(29) CATNAP Tools https://www.hiv.lanl.gov/components/sequence/HIV/neutralization/ (accessed Mar 10, 2020).

(30) Ofek, G.; Tang, M.; Sambor, A.; Katinger, H.; Mascola, J. R.; Wyatt, R.; Kwong, P. D. Structure and Mechanistic Analysis of the Anti-Human Immunodeficiency Virus Type 1 Antibody 2F5 in Complex with Its Gp41 Epitope. *J. Virol.* **2004**, *78* (19), 10724–10737. https://doi.org/10.1128/JVI.78.19.10724-10737.2004.

(31) Kwon, Y. D.; Georgiev, I. S.; Ofek, G.; Zhang, B.; Asokan, M.; Bailer, R. T.; Bao, A.; Caruso, W.; Chen, X.; Choe, M.; Druz, A.; Ko, S.-Y.; Louder, M. K.; McKee, K.; O'Dell, S.; Pegu, A.; Rudicell, R. S.; Shi, W.; Wang, K.; Yang, Y.; Alger, M.; Bender, M. F.; Carlton, K.; Cooper, J. W.; Blinn, J.; Eudailey, J.; Lloyd, K.; Parks, R.; Alam, S. M.; Haynes, B. F.; Padte, N. N.; Yu, J.; Ho, D. D.; Huang, J.; Connors, M.; Schwartz, R. M.; Mascola, J. R.; Kwong, P. D. Optimization of the Solubility of HIV-1-Neutralizing Antibody 10E8 through Somatic Variation and Structure-Based Design. *J. Virol.* **2016**, *90* (13), 5899–5914. https://doi.org/10.1128/JVI.03246-15.

(32) Irimia, A.; Sarkar, A.; Stanfield, R. L.; Wilson, I. A. Crystallographic Identification of Lipid as an Integral Component of the Epitope of HIV Broadly Neutralizing Antibody 4E10. *Immunity* **2016**, *44* (1), 21–31. https://doi.org/10.1016/j.immuni.2015.12.001.

(33) Fleury, D.; Barrère, B.; Bizebard, T.; Daniels, R. S.; Skehel, J. J.; Knossow, M. A Complex of Influenza Hemagglutinin with a Neutralizing Antibody That Binds Outside the Virus Receptor Binding Site. *Nat. Struct. Biol.* **1999**, *6* (6), 530–534. https://doi.org/10.1038/9299.

(34) Pejchal, R.; Gach, J. S.; Brunel, F. M.; Cardoso, R. M.; Stanfield, R. L.; Dawson, P. E.; Burton, D. R.; Zwick, M. B.; Wilson, I. A. A Conformational Switch in Human Immunodeficiency Virus Gp41 Revealed by the Structures of Overlapping Epitopes Recognized by Neutralizing Antibodies. *J. Virol.* **2009**, *83* (17), 8451–8462. https://doi.org/10.1128/JVI.00685-09.

(35) Ekiert, D. C.; Friesen, R. H. E.; Bhabha, G.; Kwaks, T.; Jongeneelen, M.; Yu, W.; Ophorst, C.; Cox, F.; Korse, H. J. W. M.; Brandenburg, B.; Vogels, R.; Brakenhoff, J. P. J.; Kompier, R.; Koldijk, M. H.; Cornelissen, L. A. H. M.; Poon, L. L. M.; Peiris, M.; Koudstaal, W.; Wilson, I. A.; Goudsmit, J. A Highly Conserved Neutralizing Epitope on Group 2 Influenza A Viruses. *Science* **2011**, *333* (6044), 843–850. https://doi.org/10.1126/science.1204839.

(36) Ochoa, W. F.; Kalko, S. G.; Mateu, M. G.; Gomes, P.; Andreu, D.; Domingo, E.; Fita, I.; Verdaguer, N. A Multiply Substituted G–H Loop from Foot-and-Mouth Disease Virus in Complex with a Neutralizing Antibody: A Role for Water Molecules. *J. Gen. Virol.* **2000**, *81* (6), 1495–1505. https://doi.org/10.1099/0022-1317-81-6-1495.

(37) Corti, D.; Voss, J.; Gamblin, S. J.; Codoni, G.; Macagno, A.; Jarrossay, D.; Vachieri, S. G.; Pinna, D.; Minola, A.; Vanzetta, F.; Silacci, C.; Fernandez-Rodriguez, B. M.; Agatic, G.; Bianchi, S.; Giacchetto-Sasselli, I.; Calder, L.; Sallusto, F.; Collins, P.; Haire, L. F.; Temperton, N.; Langedijk, J. P. M.; Skehel, J. J.; Lanzavecchia, A. A Neutralizing Antibody Selected from Plasma Cells That Binds to Group 1 and Group 2 Influenza A Hemagglutinins. *Science* **2011**, *333* (6044), 850–856. https://doi.org/10.1126/science.1205669.

# Supporting Information
# Potential Neutralizing Antibodies Discovered for Novel Corona Virus Using Machine Learning

*Rishikesh Magar [¥], Prakarsh Yadav [‡] and Amir Barati Farimani [1¥§‡]*

[¥] Department of Mechanical Engineering, Carnegie Mellon University, Pittsburgh, PA 15213
[‡] Department of Biomedical Engineering, Carnegie Mellon University, Pittsburgh, PA 15213
[§] Machine Learning Department, School of Computer Science, Carnegie Mellon University
Pittsburgh, PA 15213

---

[1] Corresponding Author, e-mail: barati@cmu.edu

**Supplementary Figure 1.** Contact Distance plot of trajectories of all the 22 mutant structures and the crystal structures 2GHW and 6NB6. The contact distances were computed using MDTraj package. Instantaneous contact distances for each timestep are shown for the course of the simulation (15 ns). The contact distance values vary within a small range for all the structures, indicating that simulations did not have any abrupt changes.

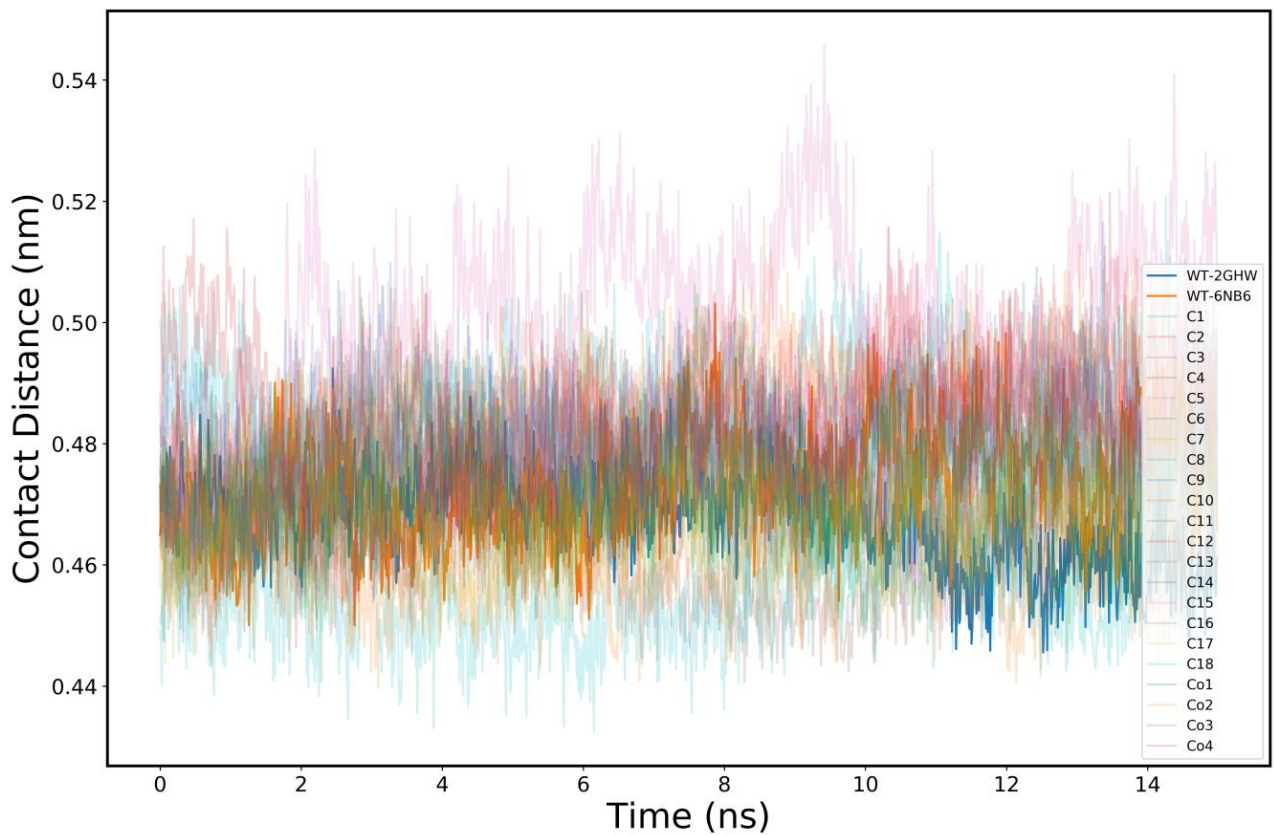

**Supplementary Figure 2.** Root Mean Square Deviation (RMSD) plot of trajectories of all the 22 structures and the crystal structures 2GHW and 6NB6. The deviations are within acceptable range indicating the strucutres were stable over the course of simulation. RMSD was averaged over the entire simultation and subsequently used for generation of Figure 4b.

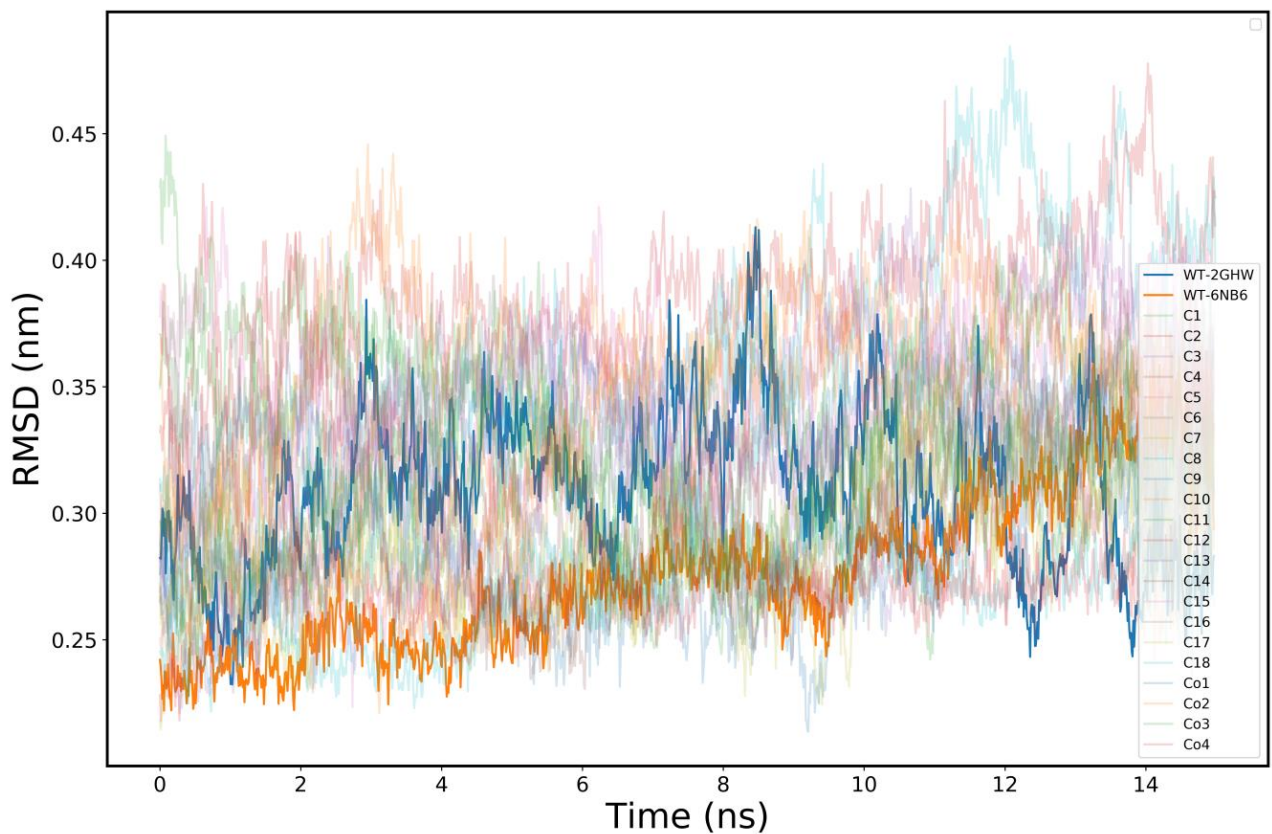

**Supplementary Figure 3.** Structure of virus antigen and antibody along with the epitope interactions in the crystal structure PDB ID: 2GHW. The viral antigen is in green and antibody is in cyan. Black lines indicate electrostatic interactions between the antigen epitope and the antibody. These interactions play a significant role in the specificity of antigen recognition by antibody.

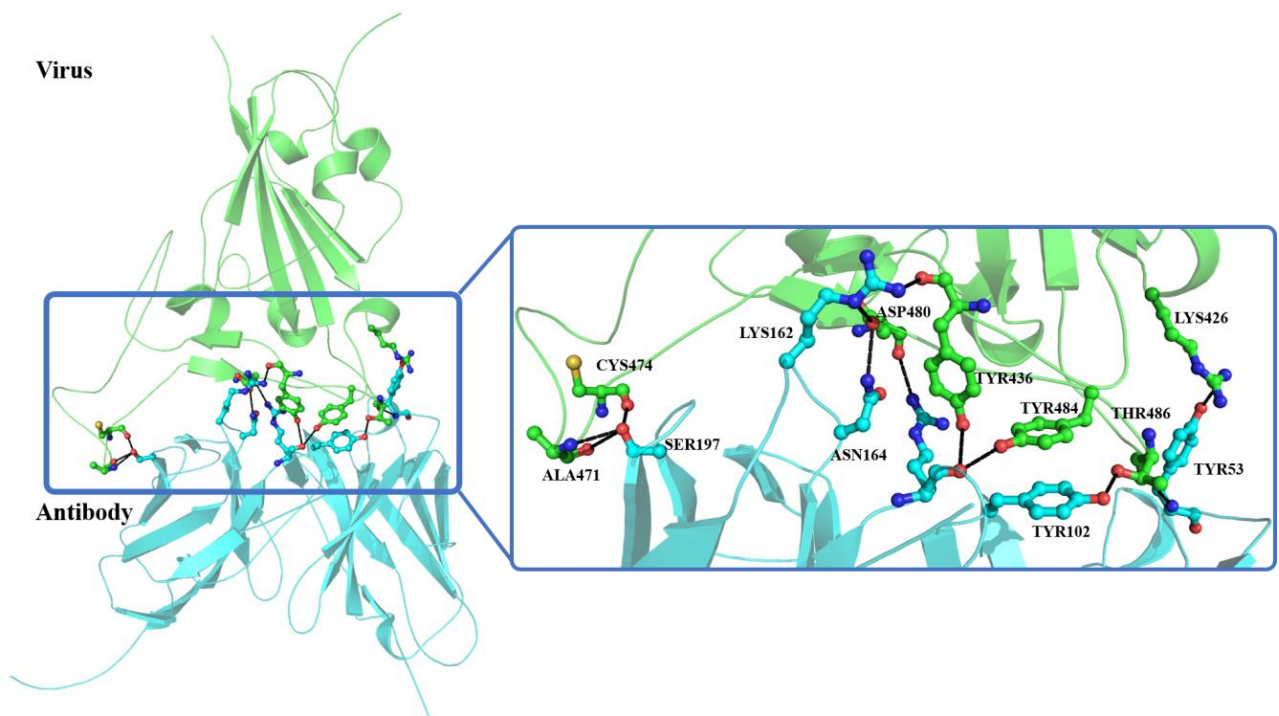

**Supplementary Figure 4.** Native contacts between the antibody and antigen in the crystal structure (PDB ID: 2GHW) shown pictographically. The amino acids above black dotted line represents the antibody (pink) and below it is the viral epitope (brown). Electrostatic interactions are represented by the whole amino acid ball-stick diagram and are connected by green dotted lines.

**Supplementary Figure 5.** Predicted contacts between the antibody and COVID-19 antigen. The COVID antigen is overlaid onto the crystal structure antigen (2GHW) to conserve the geometry and make predictions. The amino acids above black dotted line represents the antibody (pink) from PDB: 2GHW, and below is the viral epitope(brown) from COVID-19 sequence. Electrostatic interactions are represented by the whole amino acid ball-stick diagram and are connected by green dotted lines.

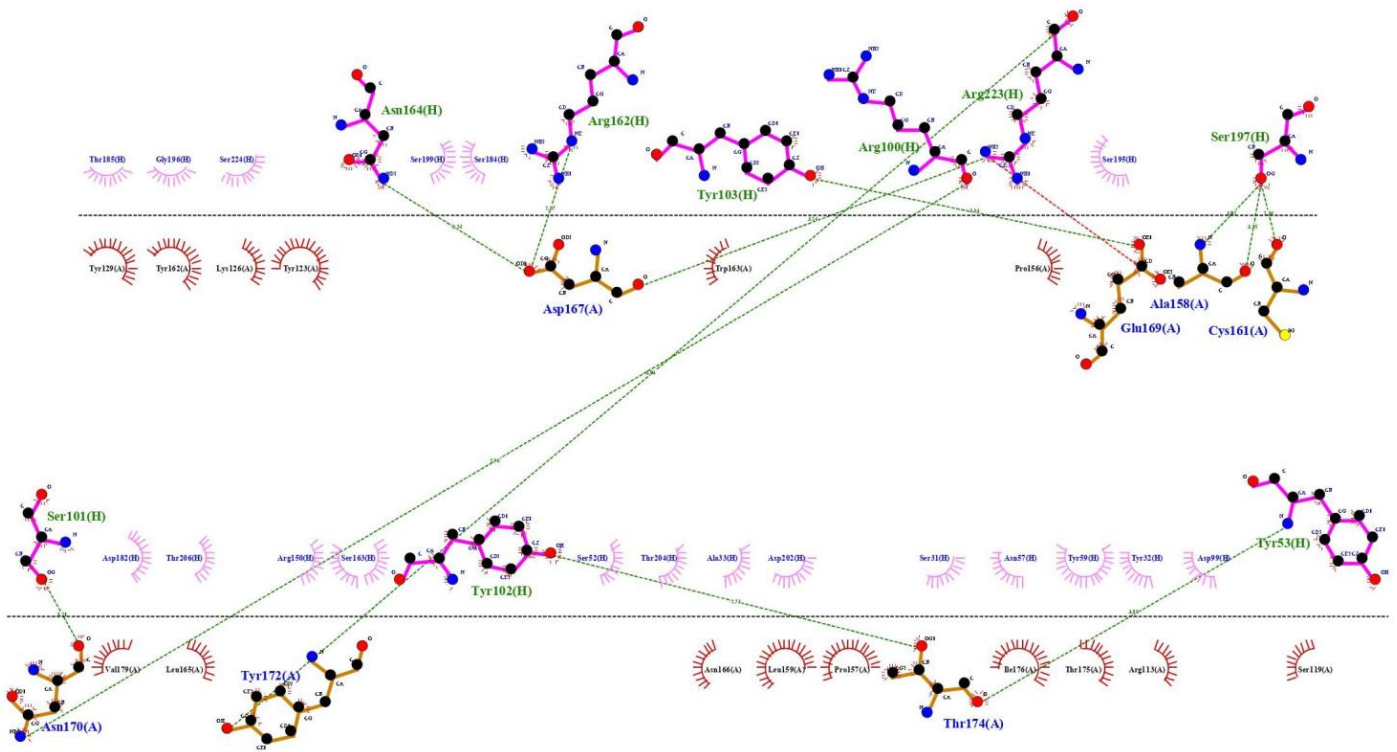

**Table S1:** The neutralizing candidates obtained before screening for stability, and after screening with ML and Bioinformatics. There are 18-point mutations in total which are predicted to have a neutralizing effect on COVID-19.

| Structure | Mutation |
|---|---|
| C1 | 2GHW – A33C |
| C2 | 2GHW – V50M |
| C3 | 2GHW – I51M |
| C4 | 2GHW – I51V |
| C5 | 2GHW – N57H |
| C6 | 2GHW – R100H |
| C7 | 2GHW – R150H |
| C8 | 2GHW – V161M |
| C9 | 2GHW – R162H |
| C10 | 2GHW – N164H |
| C11 | 2GHW – T185N |
| C12 | 2GHW – R186H |
| C13 | 2GHW – F203M |
| C14 | 2GHW – T204N |
| C15 | 2GHW – T206N |
| C16 | 2GHW – R223H |
| C17 | 6NB6 – R56H |
| C18 | 6NB6 – K58S |

**Table S2:** The possible co-mutations from 5 stable neutralizing candidates. Structures Co1 – Co4 were predicted as neutralizing, however; structure Co5 was predicted as non-neutralizing by the XGBoost model. All these co-mutant structures we MD validated for their stability.

| Structure | Mutation | Neutralization Potential |
|:---:|:---:|:---:|
| Co1 | 2GHW – I51M, R150H, T204N | 1 |
| Co2 | 2GHW – I51M, R150H | 1 |
| Co3 | 2GHW – I51M, T204N | 1 |
| Co4 | 6NB6 – R56H, K58S | 1 |
| Co5 | 2GHW – R150H, T204N | 0 |

**Supplementary Data.** Stable_str.zip. The zip file contains the PDB coordinates for the final frame of the simulation of each of the neutralizing candidate antibody. All the mutant antibodies are structurally intact and are stable.

**Supplementary Methods**

**IC$_{50}$ Value interpretation**

The IC$_{50}$ values in dataset varied from 0 to >50. For simplicity of analysis, we classified the IC$_{50}$ values into two categories. Where, IC$_{50}$ > 10 would indicate a non-neutralizing antibody and IC$_{50}$ < 10 would indicate a neutralizing antibody. Many viruses had the same viral epitope and yet had very different IC$_{50}$ values. As a result, the model had to predict two different labels for the same antibody-antigen combinations. We resolved this, by taking a majority vote among the conflicts and transformed the labels which do not concur with the majority. Employing this scheme improved the results of our model.